\documentstyle [11pt,menu97]{article}
\begin{document}
    \setlength{\baselineskip}{2.6ex}
\title{Analysis of the $pp \to pp$, $\pi d\to \pi d$, and $\pi 
d\to pp$ scattering data}
\author{Richard Arndt, Chang-Heon Oh, Igor Strakovsky, and Ron Workman\\
{\em Department of Physics, Virginia Polytechnic Institute and State 
University\\ 
Blacksburg, VA 24061-0435}}
\maketitle
\begin{abstract}
\setlength{\baselineskip}{2.6ex}
A combined analysis of the main reactions of the two-baryon system 
($pp\to pp$, $\pi d\to \pi d$ and $\pi d\to pp$) over the $\sqrt{s}$ 
interval from pion threshold to 2.4~GeV has been completed.  The 
overall phase in $\pi d\to pp$ has now been determined.  The combined 
analysis has resulted in an improved fit to the $\pi d$ elastic and 
$\pi d\to pp$ databases.
\end{abstract}

\section*{INTRODUCTION}

Below 2~GeV, in the proton laboratory kinetic energy $T_{p}$ for the 
$NN$ system, the dominant channels contributing to $NN$ inelasticity 
are $\pi d$ and $N\Delta$\cite{1}.  In Fig.~1 we display the total 
cross sections for $pp$ and $\pi d$ scattering broken into their 
components.
At these energies, it is useful to employ a multi-channel formalism in 
analyzing all existing data simultaneously.  In the present work, we 
have used the K-matrix formalism in order to unify the analysis of 
coupled reactions, $pp\to pp$\cite{2}, $\pi d\to \pi d$\cite{3}, and 
$\pi d\to pp$\cite{4}, which we have, in the past, considered 
separately.  The range of $\sqrt{s}$ was chosen to include all of our 
previous results for the pion-induced reactions ($T_{\pi} = 0-500~MeV$, 
or $T_p = 288-1290~MeV$).  

The present analysis differs from those carried out previously\cite{5} 
in a number of important respects.  We did not restrict our study to 
partial-waves containing interesting structures.  For $pp$ elastic 
scattering, all waves with $J\le 7$ were used.  Partial waves with 
$J\le 5$ were retained for both $\pi d$ elastic scattering and $\pi 
d\to pp$.  In addition, the K-matrix parameters were determined solely 
from our fits to the available databases for each separate reaction.   
No results of outside analyses or any model approaches were used as 
constraints.  As a result, the amplitudes found in our K-matrix fits 
are as ``unbiased" as those coming from the separate analyses\cite{6}. 

\section*{FORMALISM}

We have constructed a K-matrix formalism coupling the $pp$, $\pi d$ 
and $N\Delta$-channels, in order to analyze the reaction $\pi d\to pp$ 
along with elastic $pp$ and $\pi d$ scattering.  The $N\Delta$-channel 
was added to account for all channels other than $pp$ and $\pi d$.  
This choice ensured that unitarity would not be violated in our global 
fit.
As the elastic $pp$ partial-wave analysis is far superior to the
$\pi d$ elastic and $\pi d\to pp$ analyses, we have carried out fits
in which the $pp$ partial-waves were held fixed.  As described in Ref.
\cite{7}, the $pp$ amplitudes were used to fix some elements of the 
K-matrix, while the others were determined from a fit to the combined 
$\pi d$ elastic and $\pi d\to pp$ databases.

States of a given total angular momentum and parity were parameterized 
by a (4x4) K-matrix which coupled to an appropriate $N\Delta$-channel.  
Spin-mixed (2x2) $pp$-states couple to unmixed $\pi d$-states, and 
unmixed $pp$-states couple to spin-mixed (2x2) $\pi d$-states, so the 
$\pi d - pp$ system is always represented by a (3x3) matrix.  The 
associated T-matrix elements are then expanded as polynomials in the 
pion energy times appropriate phase-space factors. 

\section*{PARTIAL-WAVE AMPLITUDES}

We have fitted the amplitudes for $pp\to pp$ and the existing databases 
for $\pi d\to pp$, and $\pi d\to \pi d$, using the K-matrix formalism 
outlined in the previous Section.  The $\pi d$ elastic and $\pi d\to 
\pi d$ databases, used in this analysis, are described in Refs.\cite{3} 
and\cite{4}, and are available from the authors\cite{6}.  The overall 
$\chi^2$ for our combined analysis is actually superior to that found 
in our separate analyses (Table~1).  This is due to the improved 
parameterization scheme.  
\begin{table}[tbh]
\caption{\it Comparison of the combined analysis (C500) and our 
previous (separate) analyses:  WI96 for $pp\to pp$\protect\cite{2}, 
SM94 for $\pi d\to \pi d$ \protect\cite{3}, and SP96 for $\pi d\to 
pp$\protect\cite{4}.  The relevant energy ranges are: $T_{\pi} = 
0-500~MeV$, $T_p = 288-1290~MeV$, and $\protect\sqrt{s} = 
2015-2440~MeV$.}
\label{tbl1}
\begin{center}
\begin{tabular}{|c|c|c|}
 \hline
Reaction          & Separate      & Combined      \\
                  & $\chi^2$/Data & $\chi^2$/Data \\
 \hline
$pp\to pp$        & $17380/10496$ & $17380/10496$ \\
$\pi d\to \pi d$  &  $2745/1362$  &  $2418/1362$  \\
$\pi d\to pp$     &  $7716/4787$  &  $7570/4787$  \\
 \hline
\end{tabular}
\end{center}
\end{table}
We should emphasize that the amplitudes for $pp$ elastic scattering 
are the same as those given in Ref.\cite{2}.  (No difference was found 
in C500 using the $pp$ solution WI96, which was limited to T$_p$ = 
1600~MeV, vs our recent solution SM97, which was extended to T$_p$ = 
2500~MeV\cite{8}.)

The results for our combined/separate analyses of $\pi d$ elastic 
scattering are also qualitatively similar, up to the limit of our 
single-energy analyses. In Fig.~2a we compare the main partial-waves 
from our separate analysis (solution SM94)\cite{3} and our combined 
analysis (solution C500).  Significant differences begin to appear 
above a pion laboratory kinetic energy of 300~MeV or 2.3~GeV in 
$\sqrt{s}$.  The upper limit to our single-energy analyses is due to 
a sharp cutoff in the number of data.  This is apparent in Fig.~2 of 
Ref.\cite{3}.  Much additional data above 300~MeV will be required 
before a stable solution to 500~MeV can be expected.

A comparison of results for $\pi d\to pp$ reveals the most pronounced 
differences.  One reason for this is the overall phase which was left 
undetermined in Ref.\cite{4}.  There, we arbitrarily chose the $^3P_1S$ 
wave to be real.  In the present analysis, the overall phase has been 
determined.  Given that the overall phase is found to be about $60 
^{\circ}$ in the combined analysis, we have chosen to compare the 
partial-wave amplitudes from the separate and combined analyses in 
terms of their moduli (Fig.~2b).  As was the case for $\pi d$ elastic 
scattering, differences are most significant above approximately 2.3~GeV 
in $\sqrt{s}$.  A similar lack of data exists above this energy.

In general, we see a good agreement for the dominant amplitudes found 
in the separate and combined analyses.  Single-energy analyses were done 
in order to search for structures which may be missing from the 
energy-dependent fit.  Those corresponding to the combined fit are 
displayed in Fig.~2.  Many of the partial-wave amplitudes from C500 
show rising imaginary parts near 500~MeV (Fig.~2), a feature absent 
in the analysis of $\pi d$ elastic data alone.

\section*{SUMMARY AND CONCLUSIONS}

We have obtained new partial-wave amplitudes for $\pi d$ elastic
scattering and the reaction $\pi d\to pp$, using a K-matrix method
which utilized information from our elastic $pp$ scattering analysis.
In addition to producing amplitudes more tightly constrained by
unitarity, we have resolved the overall phase ambiguity existing
in our separate $\pi d\to pp$ analysis.

The combined analysis has resulted in a improved fit to the $\pi d$ 
elastic and $\pi d\to pp$ databases.  The most noticeable differences 
in the partial-waves appear at higher energies where the existing 
data are sparse.  It is difficult to find cases where the fit has 
been dramatically improved.  One exception is the $\pi d$ total cross 
sections above 300~MeV.  Here the combined analysis is much more 
successful in reproducing the energy dependence (Fig.~3a).  The 
excitation function of $iT_{11}$ for $\pi d\to pp$ (Fig.~3b) is also 
suggestive, though the uncertainties are large.  Much additional data 
for $\pi d$ elastic and $\pi d\to pp$ above 300~MeV will be required 
to extend combined analysis to the corresponding range of $pp$ elastic 
scattering, say to $\sqrt {s}$ = 3~GeV\cite{8}.

The present analysis has also resulted in a unified description of the 
resonancelike behavior previously noted in our separate analyses of 
$pp$\cite{2} and $\pi d$\cite{3} elastic scattering, and the reaction 
$\pi d\to pp$\cite{4}.  This behavior\cite{9} has been variously 
described as ``resonant'' (due to the creation of dibaryon resonances) 
and ``pseudo-resonant'' (due to the $N\Delta$ intermediate state).  We 
expect that our combined analysis will further constrain models based 
on these two mechanisms.  

\section*{acknowledgments}

This work was supported in part by a U.~S. Department of Energy Grant 
DE--FG02--97ER41038.

\bibliographystyle{unsrt}

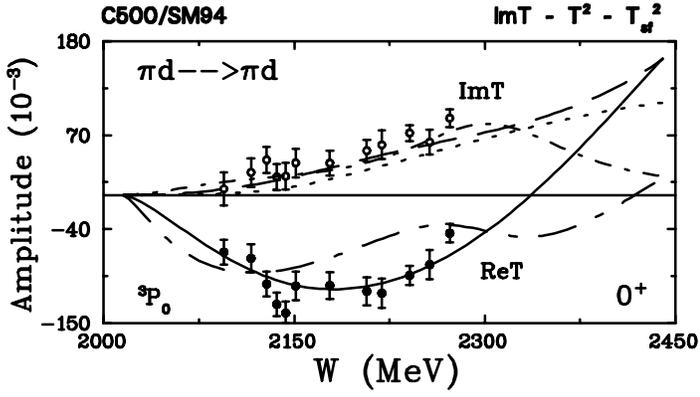
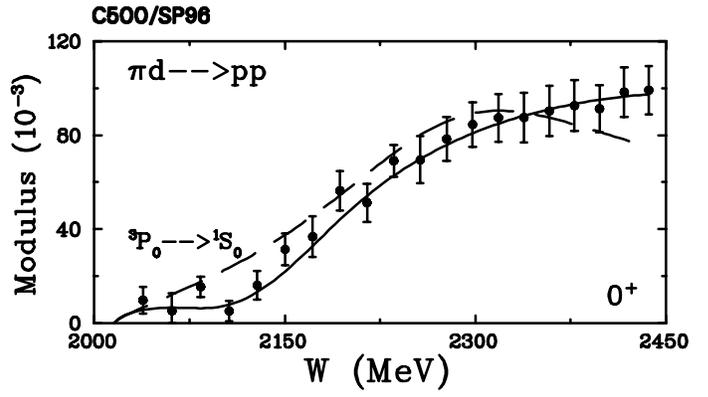
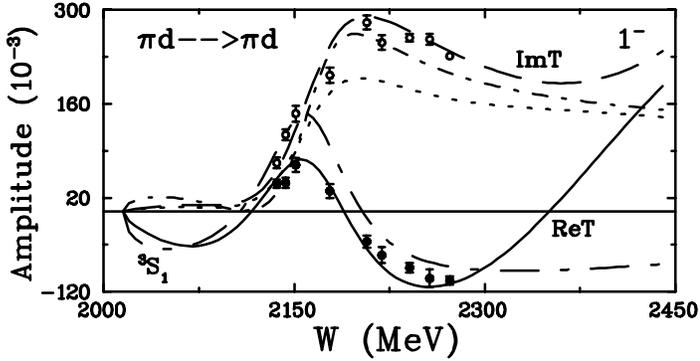
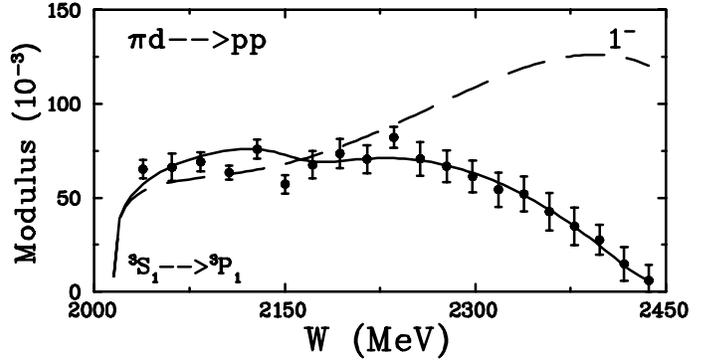
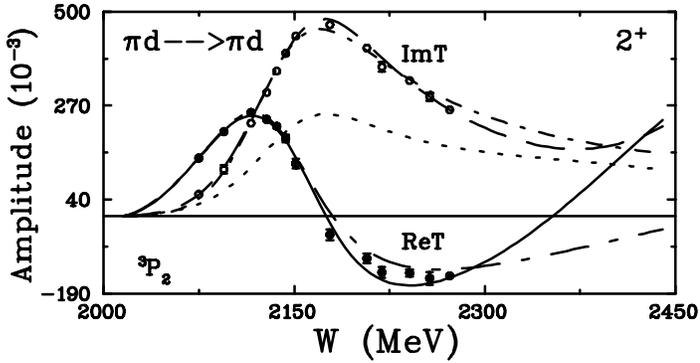
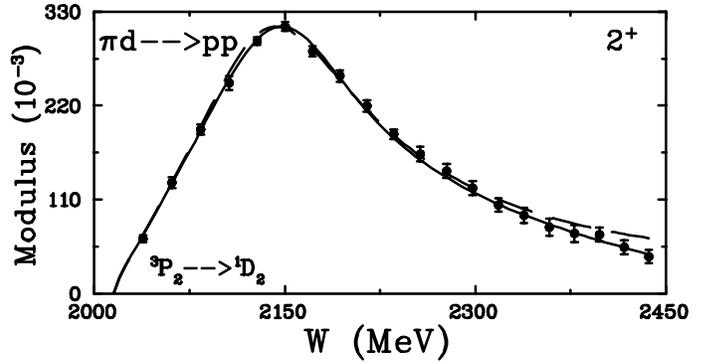
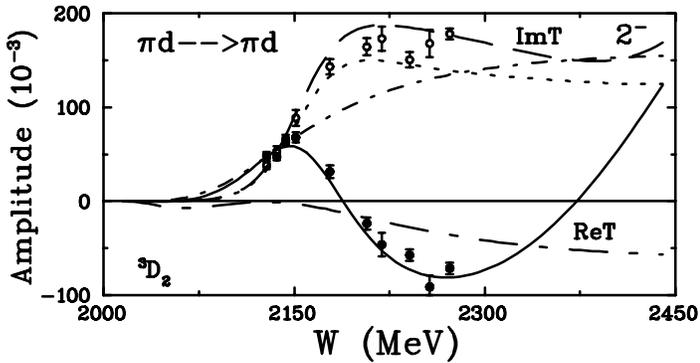
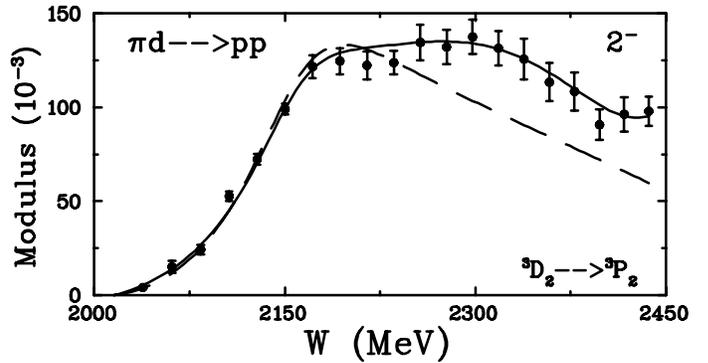
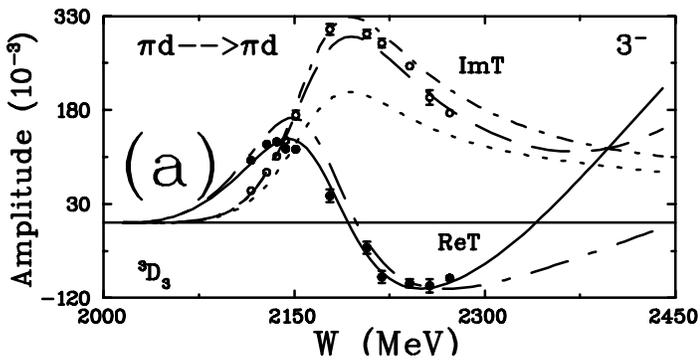
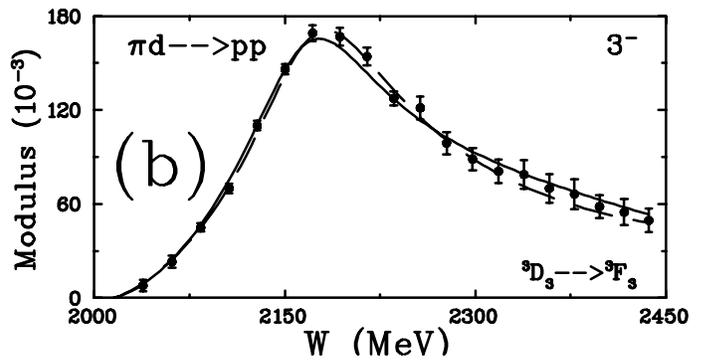
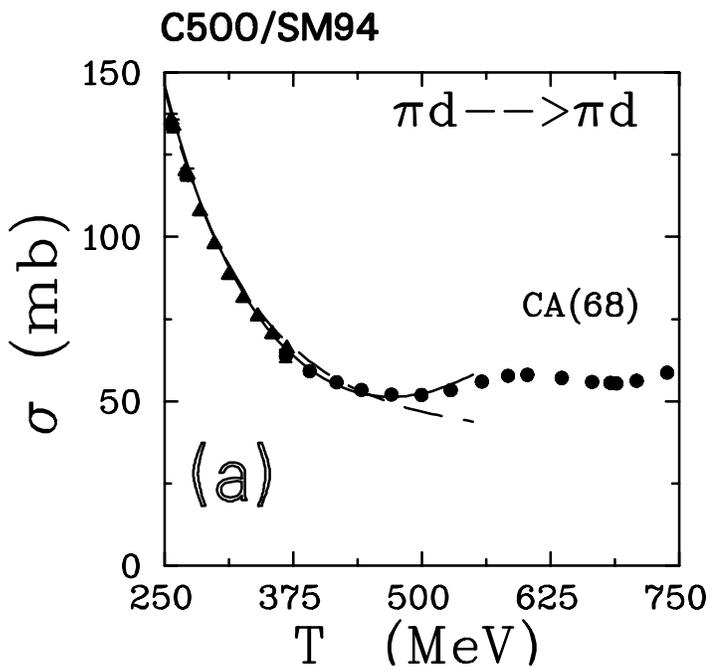
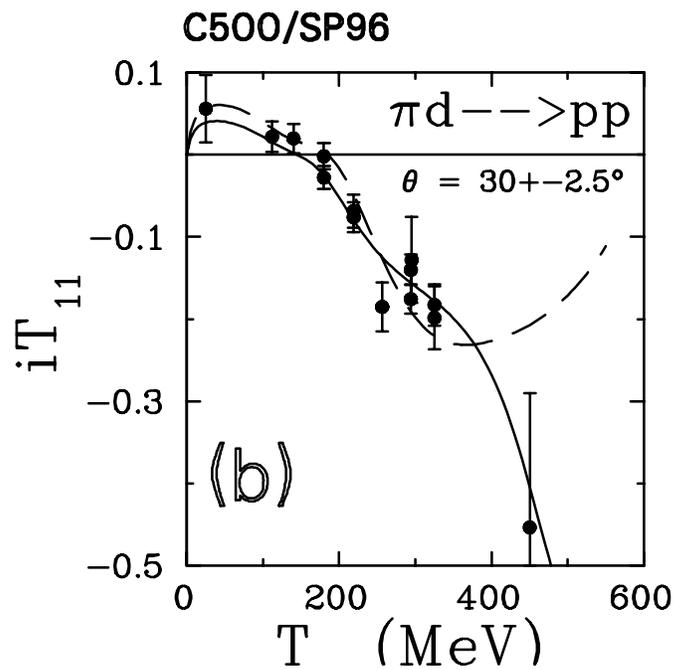
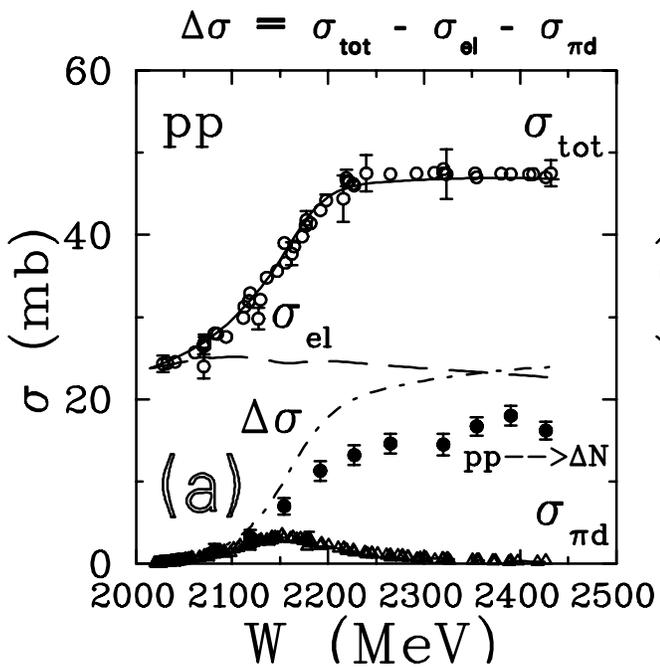
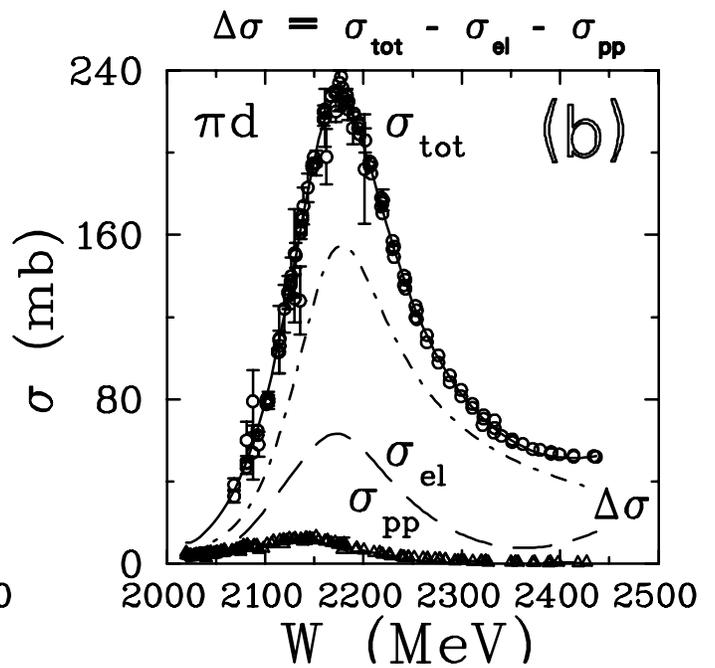
\end{document}